\documentclass[a4paper,useAMS,usenatbib]{mnras}
\usepackage{graphics,epsfig,psfig}
\hypersetup{colorlinks=true,allcolors=magenta}
\usepackage{todonotes}
\usepackage[normalem]{ulem}
\usepackage{xcolor}
\usepackage{float}
\usepackage{booktabs}
\usepackage[style=base, singlelinecheck=off, font=small, labelfont=bf, 
justification=raggedright]{caption}
\usepackage{subfigure}
\usepackage{orcidlink}

\usepackage{graphicx}
\usepackage[]{inputenc,amssymb}
\usepackage{natbib}
\usepackage{url}
\usepackage{widetext}
\usepackage{amsmath}
\usepackage{cancel}

  \makeatletter
\def\@fnsymbol#1{\ensuremath{\ifcase#1\or \dagger\or \ddagger\or
   \mathsection\or \mathparagraph\or \|\or **\or \dagger\dagger
   \or \ddagger\ddagger \else\@ctrerr\fi}}
    \makeatother
\def \be{\begin{equation}}
\def \ee{\end{equation}}
\def \bea{\begin{eqnarray}}
\def \eea{\end{eqnarray}}

\def\apj{ApJ}
\def\mnras{MNRAS}
\def\prd{PRD}
\def\aap{A\&A}
\definecolor{webgreen}{rgb}{0,.5,0}
\definecolor{webbrown}{rgb}{.6,0,0}

\def\aap{A\&A}
\def\apj{ApJ}
\def\apjs{ApJS}
\def\apjl{ApJL}
\def\mnras{MNRAS}

\def\aj{AJ}
\def\nat{Nature}

\def\prd{Phys.Rev.D}         
\def\prl{Phys.Rev.Lett}      
\def\apss{ApSS}

\title[Redshift dependence of PISN mass scale]{The redshift dependence of black hole mass distribution: Is it reliable for standard sirens cosmology?}
\author[Mukherjee (2021)]{Suvodip Mukherjee$^{1}$\thanks{smukherjee1@perimeterinstitute.ca}\orcidlink{0000-0002-3373-5236}
\\
$^{1}$ Perimeter Institute for Theoretical Physics, 31 Caroline Street N., Waterloo, Ontario, N2L 2Y5, Canada\\}
\pubyear{2021}
\begin{document}
\label{firstpage}
\pagerange{\pageref{firstpage}--\pageref{lastpage}}
\maketitle

\label{firstpage}

\begin{abstract}
An upper limit on the mass of a black hole set by the pair-instability supernovae (PISN) process can be useful in inferring the redshift of the gravitational wave (GW) sources by lifting the degeneracy between mass and redshift. However, for this technique to work, it is essential that the PISN mass-scale is redshift independent or at least has a predictable redshift dependence. We show that the observed PISN mass-scale can get smeared and the position of the PISN mass-scale is likely to exhibit a strong redshift dependence due to a combined effect from the non-zero value of the delay time between the formation of a star and the merging of two black holes and the metallicity dependence of PISN mass scale.  Due to the unknown form of the delay-time distribution, the redshift dependence of the PISN mass cut-off of the binary black holes (BBHs) cannot be well characterized and will exhibit a large variation with the change in redshift. As a result, the use of a fixed PISN mass scale to infer the redshift of the BBHs from the observed masses will be systematically biased. Though this uncertainty is not severe for the third observation run conducted by the LIGO-Virgo-KAGRA collaboration, in the future this uncertainty will cause a systematic error in the redshift inferred from the PISN mass scale. The corresponding systematic error will be a bottleneck in achieving a few percent precision measurements of the cosmological parameters using this method in the future. 
\end{abstract}

\begin{keywords} 
gravitational waves, black hole mergers, cosmology: miscellaneous
\end{keywords}
\section{Introduction}
Inference of the cosmological parameters from the gravitational wave (GW) sources is one of the key science goals of the currently ongoing network of GW detectors \citep{KAGRA:2013pob} such as LIGO \citep{LIGOScientific:2014pky}, Virgo \citep{Acernese_2014}, and for the upcoming GW detectors such as KAGRA \citep{KAGRA:2020tym}, LIGO-India \citep{Unnikrishnan:2013qwa}, LISA \citep{2017arXiv170200786A}, Cosmic Explorer  \citep{Reitze:2019iox,2019CQGra..36v5002H}, and Einstein Telescope \citep{Punturo:2010zz}, as it can provide accurate measurement to the luminosity distance of the GW sources within the framework of the general theory of relativity, and without invoking any additional distance calibration. This was shown for the first time in the seminal work by \citet{Schutz}, which justifies the reason for calling GW sources the standard sirens.  
However one of the essential requirements for making robust measurements of the cosmological parameters from standard sirens is to be able to make an independent and accurate inference of the redshift of these sources. Standard sirens, though excellent distance tracers, cannot provide the redshifts to the sources independently, unless one can break the degeneracy between mass and redshift, by using a known mass scale. For binary neutron star (BNS) sources, one can use the tidal deformability to break the mass-redshift degeneracy \citep{PhysRevD.85.023535, Messenger:2011gi}. On the other hand for the binary black holes (BBHs), the existence of a maximum mass of the black holes (BHs), can be used to infer the redshift to the sources \citep{Farr:2019twy, You:2020wju, Mastrogiovanni:2021wsd}. From the theory of stellar models of BH formations, it is predicted that there is a maximum mass of the BHs due to the pair-instability supernovae process (PISN) \citep{Bethe:1998bn, 1998A&A...332..173P,2002ApJ...567..532H,2002ApJ...572..407B}, and the maximum mass of the BH is expected to be between $40-50$ M$_\odot$ \citep{2019ApJ...887...53F, Renzo:2020rzx}. A recent study has also shown \citep{2019ApJ...887...53F}, that the variation of the maximum mass-scale changes by only about $10\%$ due to a large variation in the stellar metallicity and stellar winds. As a result, if the PISN mass-scale is robust within $10\%$ accuracy, it can be used to break the mass-redshift degeneracy and can be used to infer the cosmological parameters which affect the cosmic expansion history. Even in the absence of an accurate theoretical prediction of the maximum mass value, one can do a joint estimation of the cosmological parameters and the PISN mass-scale \citep{Farr:2019twy, You:2020wju,  Mastrogiovanni:2021wsd}. So, this particular method can be extremely powerful also for the third-generation GW detectors such as Cosmic Explorer \citep{Reitze:2019iox,2019CQGra..36v5002H}, and Einstein Telescope \citep{Punturo:2010zz} to infer the cosmic expansion history to a high redshift $z\approx 50$, which is unlikely from any electromagnetic probes in the same time scale.  

However, for the usage of the PISN mass-scale for the cosmological purpose, the mass-scale must be redshift independent, or at least the redshift dependence should be predictable. If the PISN mass scale evolves with cosmological redshift, then our ignorance of the redshift dependence on the PISN mass scale, will lead to a systematic error in the measurement of the true cosmological redshift and hence will affect the inference of cosmological redshift that will use this redshift. Several previous studies assumed that the PISN mass-scale is redshift independent or will exhibit mild redshift dependence based on the theoretical studies \citep{2019ApJ...887...53F, Renzo:2020rzx}, and as the stellar metallicity at low redshift ($z<2$) varies not very significantly. In this work, we scrutinize whether the assumption that the PISN mass-scale is redshift independent is valid and show whether we can use it reliably to infer the true redshift to the binary BHs. Recent studies have explored the variation of the mass distribution with redshift for a fixed model of cosmology from GWTC-2 \citep{2021ApJ...912...98F} and GWTC-3 \citep{LIGOScientific:2021psn} and did not find any statistically significant deviation from the redshift independent scenario.    

We show that even though the variation of the PISN mass-scale is below $10\%$ across a large variation in the metallicity range ($Z \in \{10^{-3}- 10^{-5}\})$ \citep{2019ApJ...887...53F}, the delay time between the formation of the stars and merger of the BBHs will play a key role in the observed mass distribution of the BBHs. The observed mass distribution of the BBHs at a redshift will arise from the BHs formed over a wide range of cosmological redshifts due to a non-zero value of the delay time between the formation of star and merger of BHs. As a result, the mass distribution of the BBHs at any redshift will come from a vast range of redshifts over which the stellar metallicity can vary significantly. In this paper, we show the impact of the observed mass distribution of BBHs on the inferred redshift if a fixed value of the PISN mass scale M$_{\rm PISN}$ is assumed. 

The paper is organized as follows, in Sec. \ref{formalism} we introduce the idea behind the redshift dependence of the PISN mass scale and show its impact on the GW source population in Sec. \ref{redi}. In Sec. \ref{inference} we show the inference of the PISN mass scale from the mock GW samples for the network of LIGO-Virgo-KAGRA (LVK) detectors and Cosmic Explorer (CE) and in Sec. \ref{cosmo} we discuss its impact on the GW population and the cosmological parameters. Finally in Sec. \ref{conc}, we discuss the conclusion and the future prospects.

\begin{figure*}
    \centering
    \includegraphics[trim={0.cm 0.cm 0.cm 0.cm},clip,width=1.\textwidth]{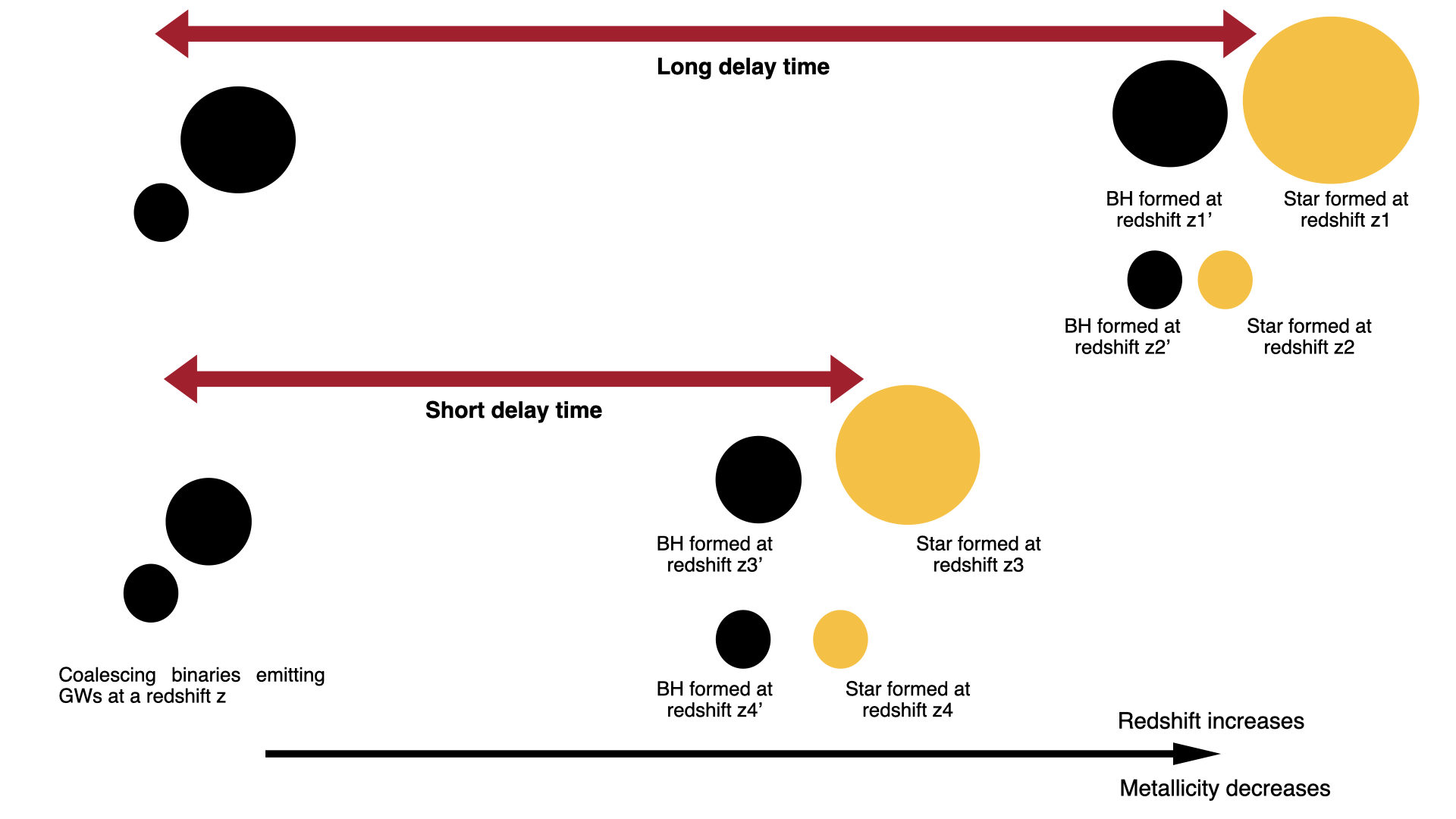}
     \captionof{figure}{A schematic diagram showing the source of variation in the PISN mass of the detected GW sources merging at a fixed redshift z. BHs formed at different redshifts can merge at the same redshift due to a delay time distribution with a non-zero value of the minimum delay time. The sources at high redshift can have lower metallicity and can have a larger value of the PISN mass scale than the BHs formed at low redshift. So, the PISN mass of the detected GW sources will exhibit a variation with redshift.}
    \label{fig:schematic}
\end{figure*}

\section{Lower edge of the PISN mass-gap  and its redshift dependence}\label{formalism}
The gap in the mass distribution of stellar-mass BHs is expected due to the mass loss of the heavy stars due to the PISN process \citep{1964ApJS....9..201F, 1967ApJ...148..803R, 1968Ap&SS...2...96F, 1984ApJ...280..825B, RevModPhys.74.1015, Talbot:2018cva, 2019ApJ...882...36M, 2019ApJ...882..121S}. The lower limit of the mass gap is expected to be around $45$ M$_\odot$, which is set by the mass loss during the PISN process \citep{2019ApJ...887...53F, Renzo:2020rzx}. The value of the lower limit of the PISN mass is shown to vary by only about $7\%$ for the variation in the stellar metallicity $Z$ from $10^{-5}$ to $3\times10^{-3}$, as the mechanism of wind-loss depends strongly on the stellar metallicity \citep{2001A&A...369..574V,2007A&A...473..603M}. Though multiple other parameters such as the nuclear rates can affect the PISN mass-scale $M_{\rm PISN}= 45$ M$_\odot$ by about $35\%$, this parameter is not expected to evolve with cosmological redshift and hence cannot cause redshift evolution of the PISN mass-scale. So, the stability of the lower edge of the PISN mass gap can be a robust feature for cosmology. 

However, whether the PISN mass-scale can be used for reliably inferring the cosmological parameter depends on whether the observed mass distribution is redshift independent or not, or at least whether the redshift dependence of the PISN mass-scale can be well characterized. In this work, we scrutinize the possible redshift dependence of the observed mass distribution of BBHs even in the scenario when the PISN mass scale is well predicted by theoretical studies. Though in reality, there may be uncertainties in the theoretical understanding as well. As a result, the uncertainty that we are exploring in this paper should be considered as only a lower bound. There can be additional errors on top of this as well. 

We model the metallicity dependence of the PISN mass cut-off by a relation
\begin{equation}\label{met-p}
    M_{\rm PISN} (Z)= M_{\rm PISN} (Z_*) - \alpha \log_{10}(Z/Z_*),
\end{equation}
where a weak dependence on the metallicity can be captured as a logarithmic correction with a free parameter $\alpha$. For $\alpha=1.5$ at $Z_*= 10^{-4}$ and $M_{\rm PISN}(Z_*)=$ 45 M$_\odot$, we can capture the variation in the PISN mass scale shown in \citep{2019ApJ...887...53F} due to the change in  metallicity over the range $Z \in [10^{-5}, 5\times 10^{-3}]$.  Along with the evolution of the metallicity, the PISN mass scale can also vary due to changes in the fundamental physics such as the reaction rates \citep{2022ApJ...924...39M}. 

The stellar metallicity in the Universe evolves with redshift \citep{2010MNRAS.408.2115M, 2012A&A...539A.136S,2012ApJ...753...16K,2013MNRAS.430.2891D, Madau2014}. The metallicity at a high redshift ($z>2$) is much smaller in comparison to the low redshift Universe $z<2$. The first-generation stars contaminate the interstellar medium and cause a chemical evolution of the Universe. We can treat the metallicity evolution with redshift by a relation 
\begin{equation}
   \log_{10}(Z(z))= \gamma z +\zeta,  
\end{equation}
where $\gamma$ captures the redshift dependence and $\zeta$ captures the metallicity value at $z=0$ \citep{2010MNRAS.408.2115M,Madau2014}.  This relation captures the metallicity of the parent star or the gas cloud from which a star has formed. It is written to express only a mean evolution of the metallicity. Along with the mean metallicity evolution of the Universe, there is going to be a scatter in the metallicity depending on the galaxy properties. Such a source of uncertainty brings additional stochasticity to the metallicity relation. Currently, a limited number of observations \citep{2008MNRAS.383.1439G,2010MNRAS.408.2115M,2012ApJ...753...16K} are available to explore the environment dependence of the metallicity, and most of our current understandings are based on simulations\citep{2016ApJ...822..107G,2019MNRAS.484.5587T}. These studies show that the overall median metallicity dependence of the galaxies at different can be explained by power form \citep{1999ApJ...522..604P,2007ApJ...670..584Y,2019MNRAS.484.5587T}. Several studies of GW merger rates and mass distribution are performed \citep{2002ApJ...572..407B,2012ApJ...759...52D, Dominik:2014yma, Mapelli:2017hqk,2018MNRAS.474.2959G, Toffano:2019ekp, vanSon:2021zpk} which are motivated by these studies and show the black hole mass distribution can exhibit a redshift dependence. The existence of any stochasticity in the galaxy metallicity distribution will also influence the mass distribution but is currently not well known. However, as the relation given in Eq. \eqref{met-p} is in terms of the logarithm of metallicity, so the impact of fluctuation around the median value depending on the individual galaxy properties is going a small (logarithmic) change. As we are unable to measure the host of the BBH due to a large sky localization error of the BBH, we cannot directly associate the properties of galaxies with BBH source properties. So, we can only infer an ensemble average mass distribution from the GW data and the additional stochasticity (which will depend on the host properties) will appear as an additional uncertainty in the measurement of M$_{\rm PISN}$. As a result, we consider a median distribution of galaxy metallicity and the dependence of M$_{\rm PISN}$ on it.
 
The observed mass distribution of the BBHs at a redshift $z_m$ is going to contribute from the BHs which have formed at an earlier redshift $z<z_m$ due to a non-zero value of the delay time $t_d$. The formation time of the individual companion objects will be at a different redshift. As a result, the metallicity dependence of the star corresponding to that will play a role. So, for a probability distribution $P(t_d)$ of the delay time, there is going to be mixing between the BHs forming at different redshifts. The corresponding window function of masses of the BBHs mergers can be written as
\begin{equation}
    \mathcal{W}(m(z_m))= \mathcal{N}\int_{z_m}^{\infty} P_{t_d}(z_m,z')\mathcal{W}_s(m(z')) dz',
\end{equation}
where $\mathcal{N}$ is a factor to normalize the window function. The term $P_{t_d}(z_m,z')$ denotes the probability for a BBH source to the merger at a redshift $z_m$ formed at a redshift $z$ with a delay time $t_d$, and the term $\mathcal{W}_s(m(z'))$ denotes the window function of the BH masses at a redshift $z'$ with the mass $m$ denoting the mass of the individual objects in their source frame. The convolution of the probability distribution of the delay time and the probability distribution of the mass distribution gives the source frame mass distribution at redshift $z_m$. The mass distribution of BBHs at redshift $z'$ can be modified depending on the metallicity evolution in the Universe. We show a schematic diagram in Fig. \ref{fig:schematic}  explaining the variation in the PISN mass of the BHs merging at a fixed redshift. At a fixed redshift (denoted by z), the merging BBHs could have formed at different redshifts (z1', z2', z3', and z4') from the parent stars at redshifts (denoted by redshifts z1, z2, z3, and z4)\footnote{The time lapsed between the formation of a star and the BH can be negligible in comparison to the cosmic time scale of it to merge (which is a few hundred Myr to Gyr) for the heavier black holes that contribute to the PISN mass scale. The redshift of the two parent stars are nearly the same z2 $\sim$ z1 (and also z4 $\sim$ z3). However, the difference between the redshifts like z3$-$z1 of individual binary pairs can be large.}. So, the PISN mass scale of the BHs formed at different redshifts will be different for the heavier BH. The probability distribution of the delay time with a non-zero minimum value leads to the mixing of BHs formed at a different cosmic time to merge with different PISN masses. As a result, BBHs will not have a fixed PISN mass scale and can show a large variation even if the variation due to a change in the metallicity of the PISN mass is small. The signature of the delay time distribution on the BHs mass distribution can arise in all those scenarios of binary formation for which the delay time distribution can go beyond a few hundreds of Myrs. Only for the scenarios where BBHs are merging very fast (less than a few tens on Myrs), the masses of the BBHs may not show a significant difference, if the metallicity of host galaxies are similar.

The delay-time distribution and the corresponding value of the minimum delay time are not well known. Several theoretical studies are made to understand the delay time distribution and its relation with the BBH formation channels  \citep{2010ApJ...716..615O,2010MNRAS.402..371B, 2012ApJ...759...52D, Dominik:2014yma, 2016MNRAS.458.2634M, Lamberts:2016txh, 2018MNRAS.474.4997C, Vitale:2018yhm, Elbert:2017sbr, Eldridge:2018nop, 2020ApJ...896L..32C,  Buisson:2020hoq,Santoliquido:2020axb, Safarzadeh:2020qru, vanSon:2021zpk}. For a uniform in the log-space distribution of the separation between the binaries, there is going to be a power-law form of the delay time distribution with $P(t_d) \propto t_d^{-\kappa}$ with the value of $\kappa=1$ for $t_d>t_d^{\rm min}$, here $t_d^{\rm min}$ denotes the minimum value of the delay time. For $t_d
<t_d^{\rm min}$, the probability distribution is zero. However, studies have shown that the probability distribution of the delay time can have a different functional form, apart from the unknown value of the minimum delay time denoted by t$^{\rm min}_{d}$. Recently, constraints on the delay time are obtained from the individual events \citep{Fishbach:2021mhp} and the stochastic GW background \citep{Mukherjee:2021ags}. In the future, it will be possible to measure this quantity to much higher precision by correlating the astrophysical properties of the GW sources with the star formation and metallicity properties of the host galaxies, which can be inferred from different emission lines signatures \citep{Mukherjee:2021bmw}. Gravitational lensing of GW can also provide an independent way to measure the delay time distribution from the current generation detectors \citep{Mukherjee:2021qam}.   
\begin{figure*}
    \centering
    \includegraphics[trim={0.cm 0.cm 0.cm 0.cm},clip,width=0.45\textwidth]{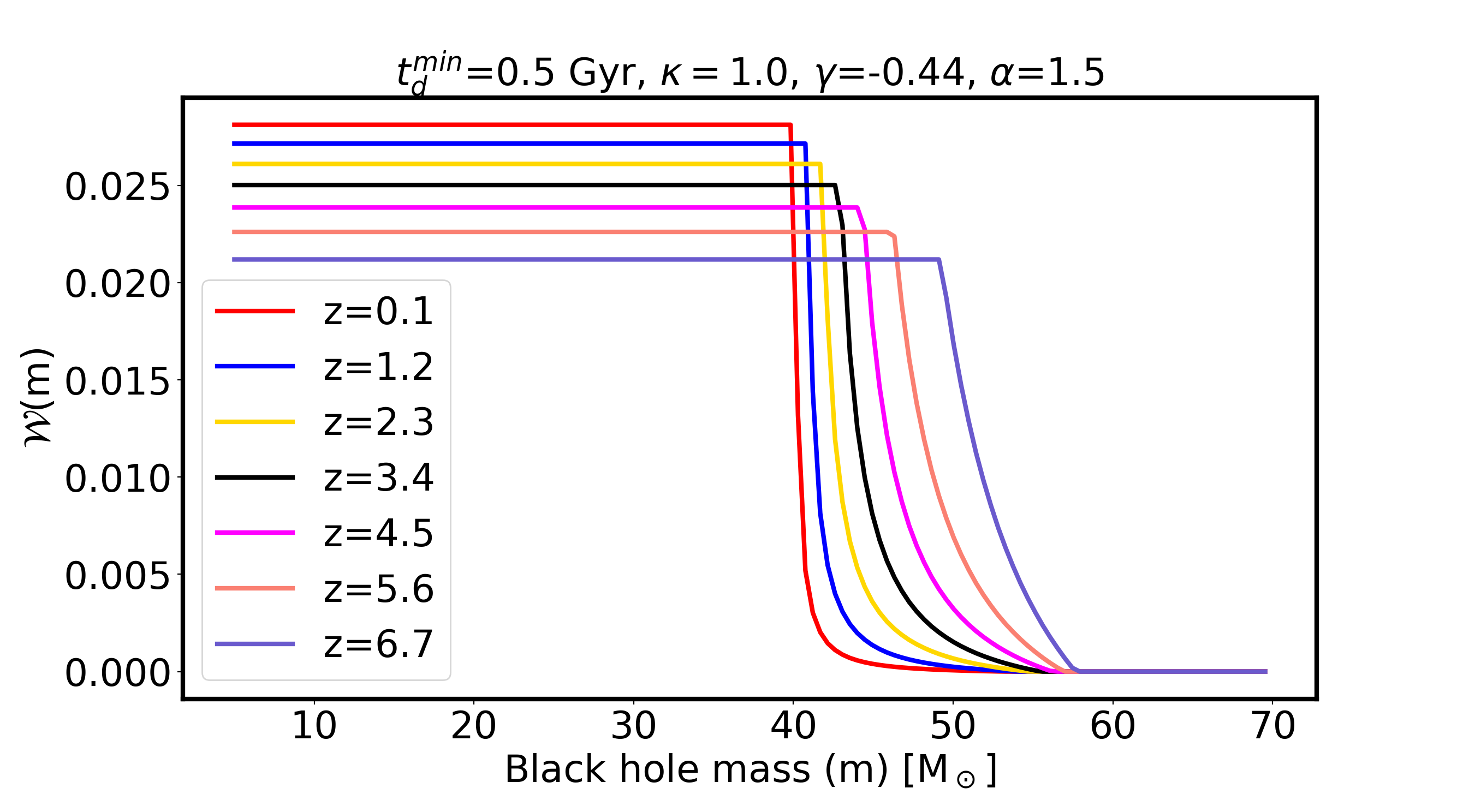}
    \includegraphics[trim={0.cm 0.cm 0.cm 0.cm},clip,width=0.45\textwidth]{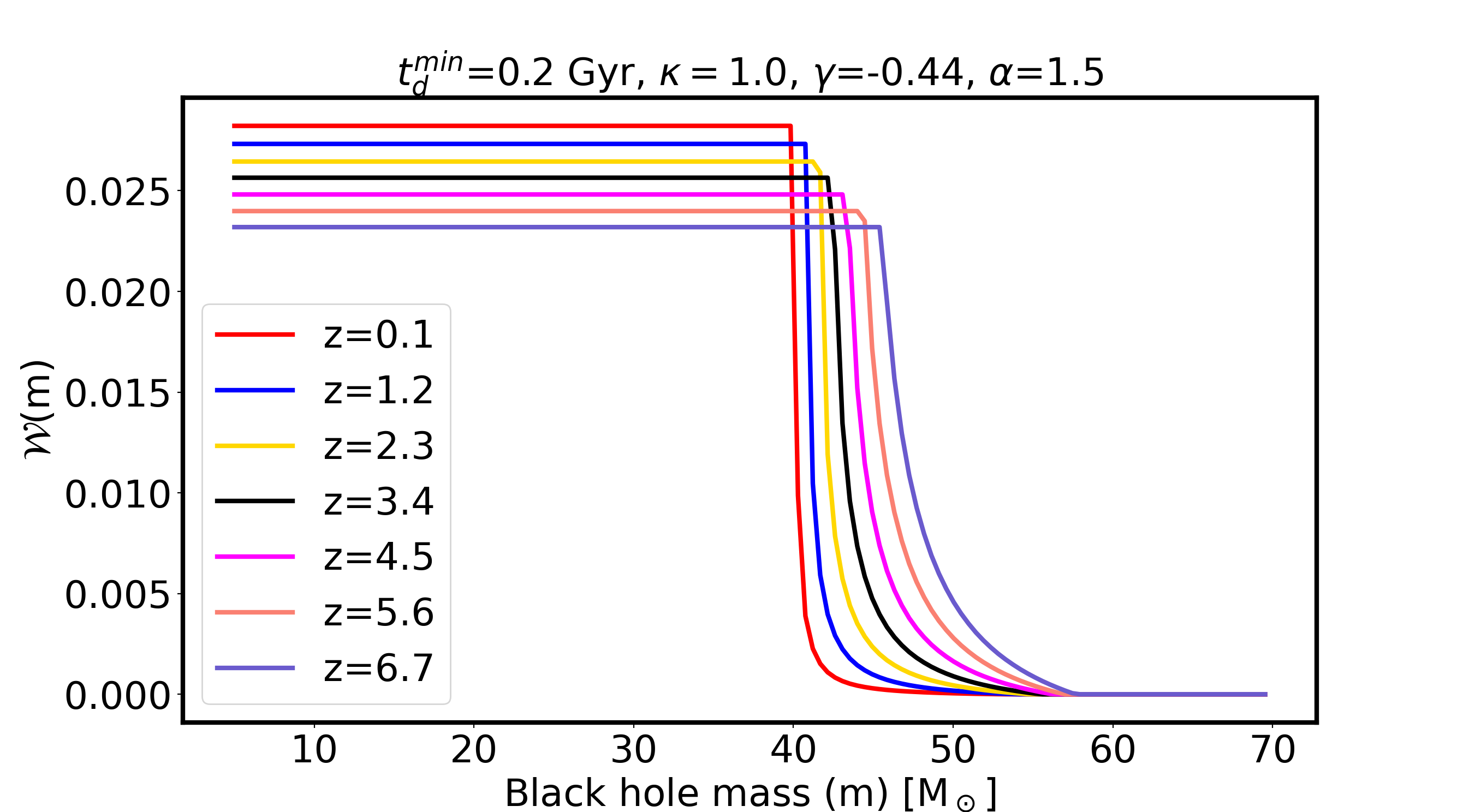}
     \includegraphics[trim={0.cm 0.cm 0.cm 0.cm},clip,width=0.45\textwidth]{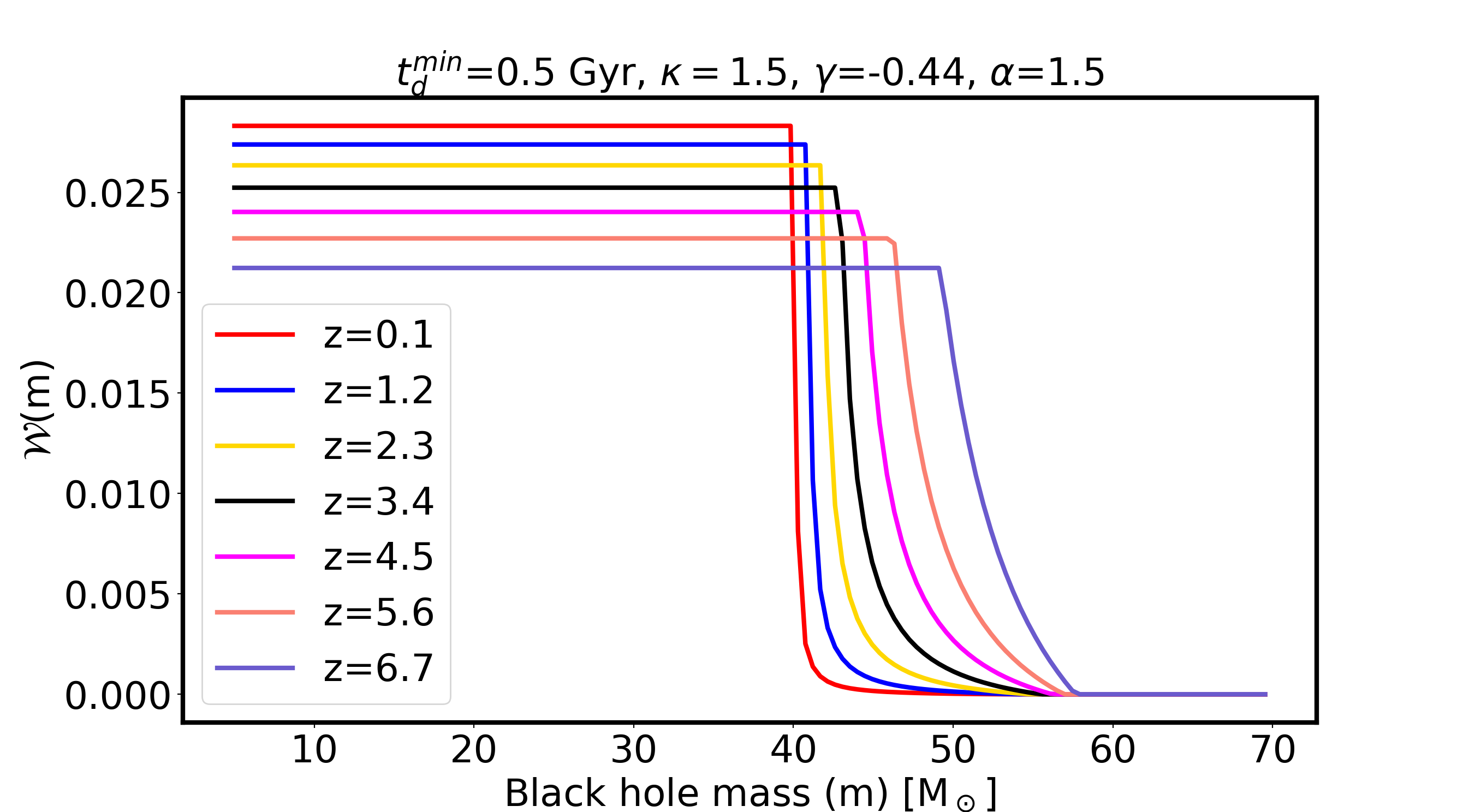}
    \includegraphics[trim={0.cm 0.cm 0.cm 0.cm},clip,width=0.45\textwidth]{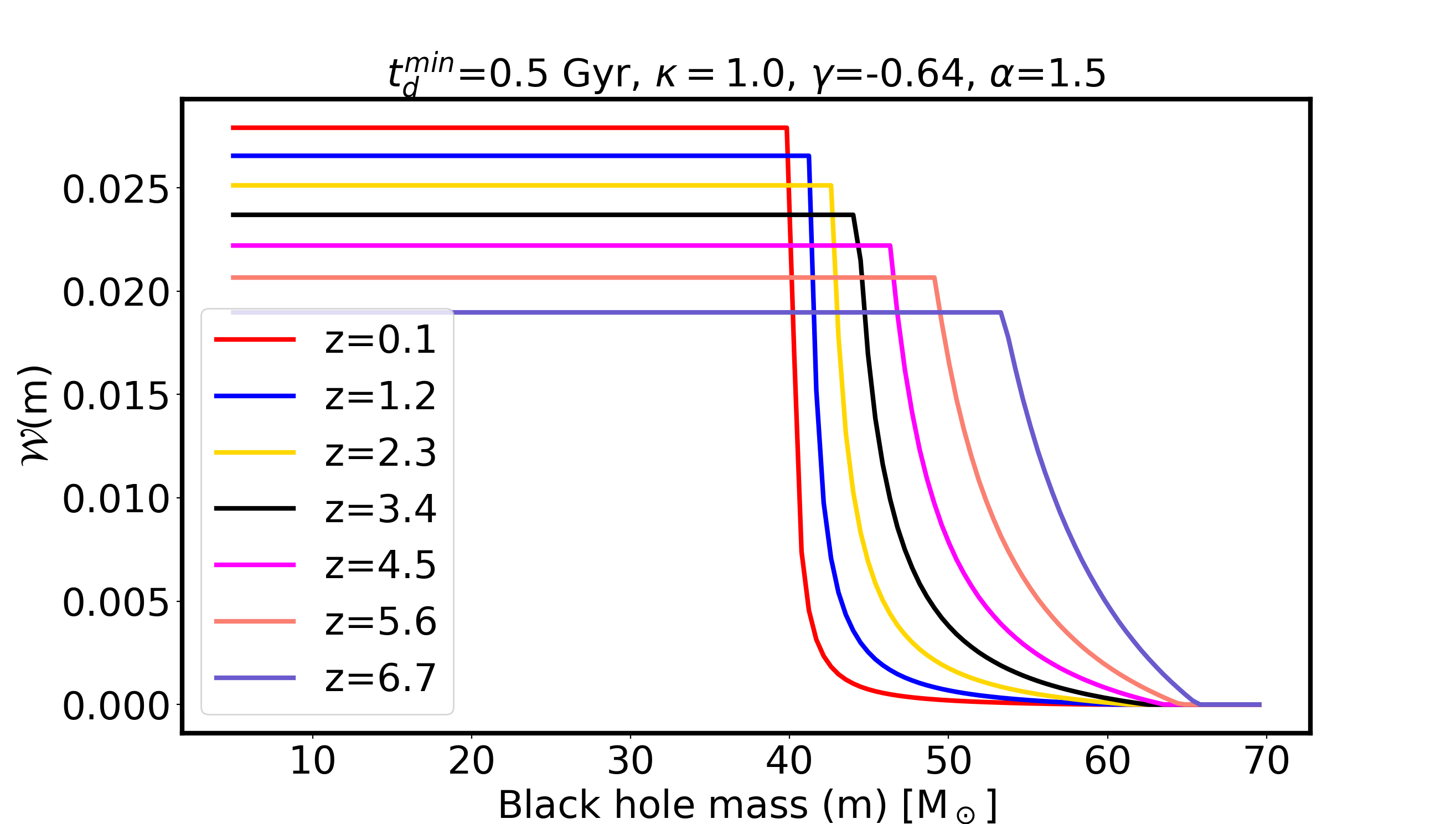}
    \includegraphics[trim={0.cm 0.cm 0.cm 0.cm},clip,width=0.45\textwidth]{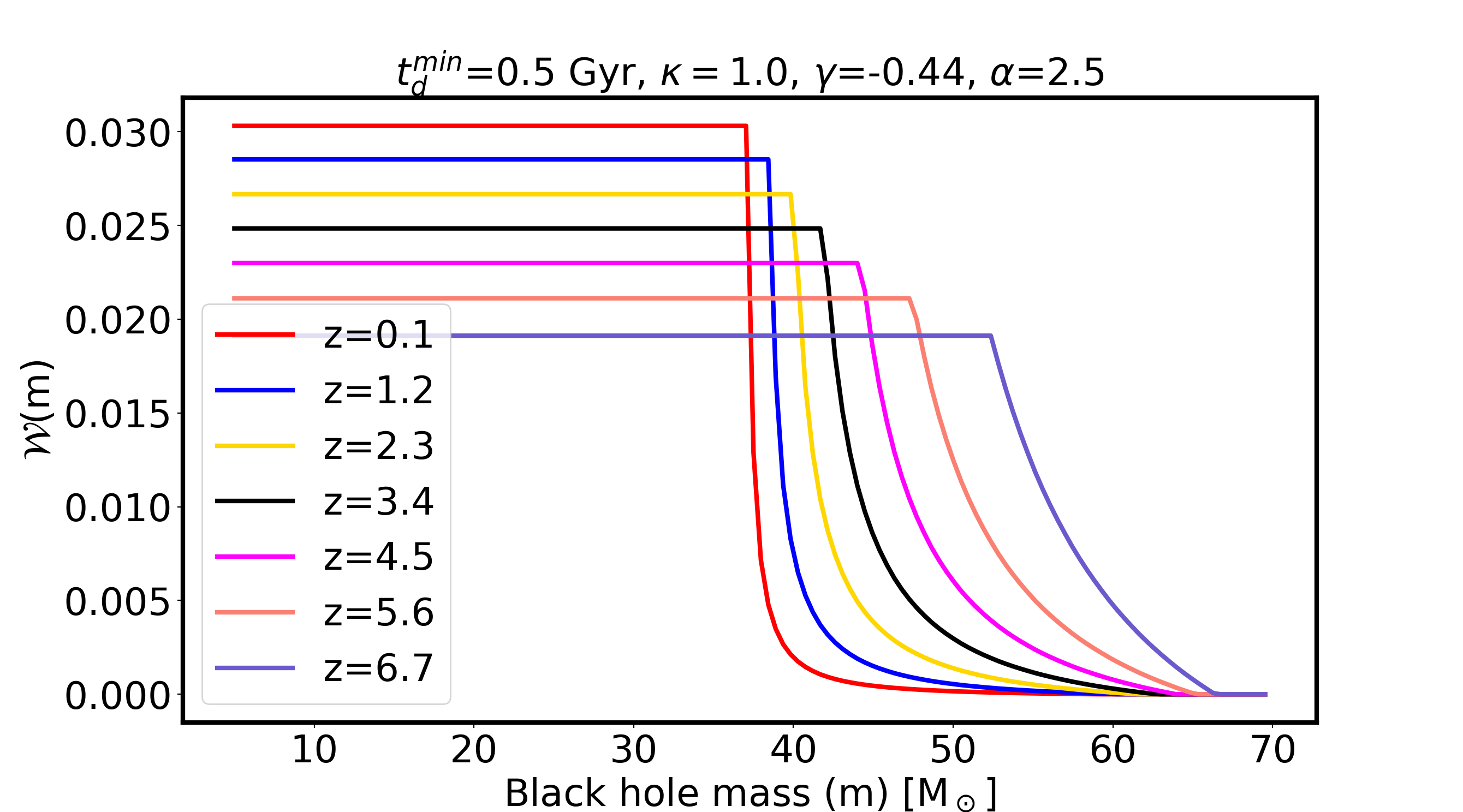}
    \caption{We show the variation in the window function for different redshifts by varying the stellar metallicity and delay time distribution.}
    \label{fig:window}
\end{figure*}

For large delay time scenarios, the component mass of the BHs can arise from a very high redshift. At high redshifts, the PISN mass-scale is going to be different and those sources contribute to the BBHs merger. For a probability distribution of the delay time $P(t_d) = t_d^{-\kappa}$ for $t_d\geq t^{\rm min}_d$, we will witness a mixing of the BHs from different redshifts to contribute at the redshift $z_m$. The probability mass distribution is going to depend on parameters $\kappa$, $t^{\rm min}_d$, $\gamma$, $\alpha$.  We show the probability distribution of mass at different redshifts for a few different values of the parameters such as $\kappa$, $t^{\rm min}_d$, $\gamma$, $\alpha$ in Fig. \ref{fig:window}. For the values chosen for the parameters $\kappa$, $t^{\rm min}_d$, $\gamma$, and  $\alpha$, the variation in the PISN mass scale is well within the variation in the mass range explored previously \citep{2002ApJ...572..407B,Belczynski:2016obo, Mapelli:2017hqk,2018MNRAS.474.2959G,Toffano:2019ekp, 2019ApJ...887...53F}.

\begin{figure*}
    \centering
    \includegraphics[trim={0.cm 0.cm 0.cm 0.cm},clip,width=1.\textwidth]{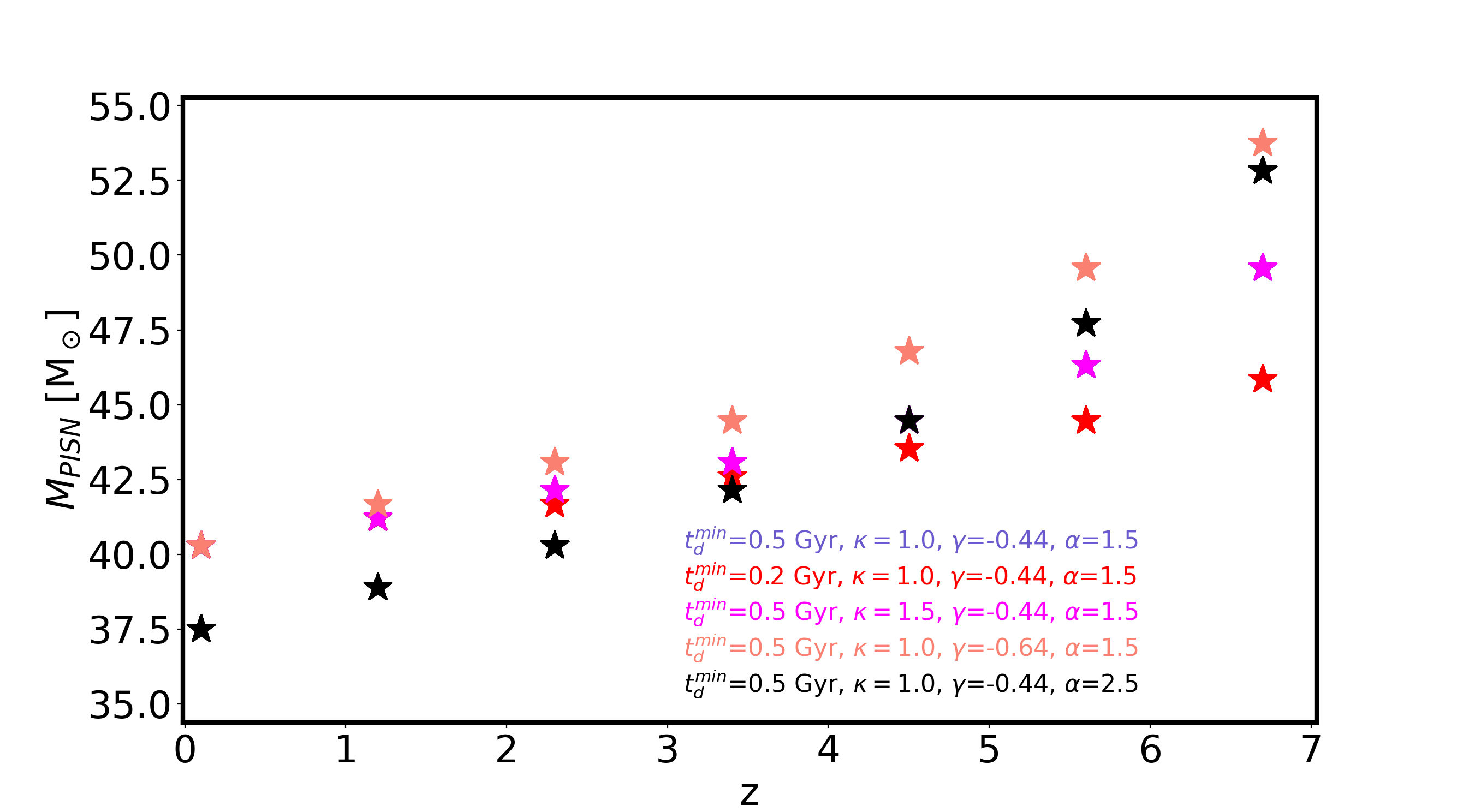}
    \includegraphics[trim={0.cm 0.cm 0.cm 0.cm},clip,width=1.\textwidth]{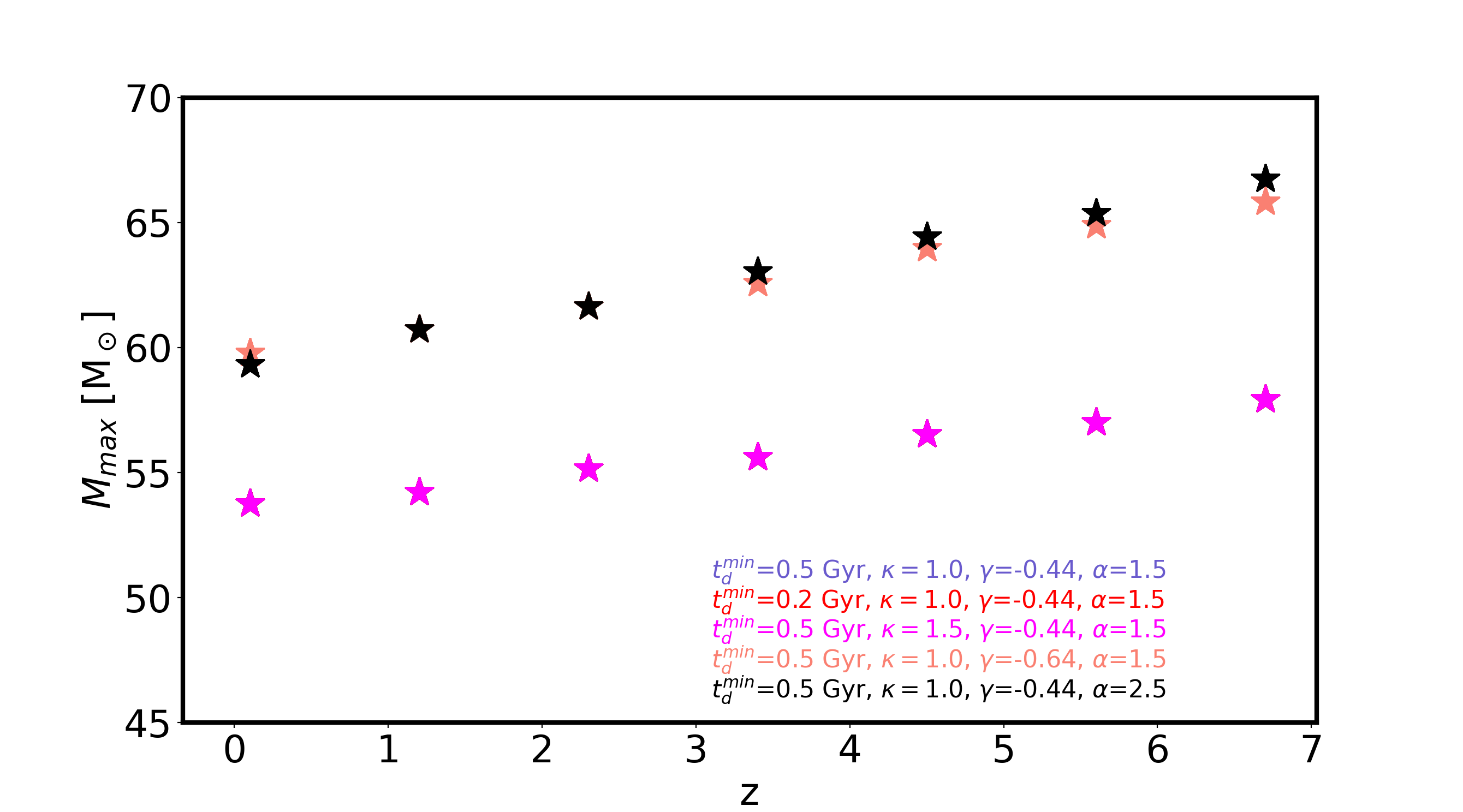}
     \caption{The variation in the PISN mass scale M$_{\rm PISN}$ (top) and maximum mass M$_{\rm max}$ (bottom) as a function of the redshift is shown for different astrophysical parameters. The blue and red markers are overlapping with the magenta markers at all the redshifts in the lower panel.}
    \label{fig:redshift}
\end{figure*}

\begin{figure}
    \centering
    \includegraphics[trim={0.cm 0.cm 0.cm 0.cm},clip,width=.45\textwidth]{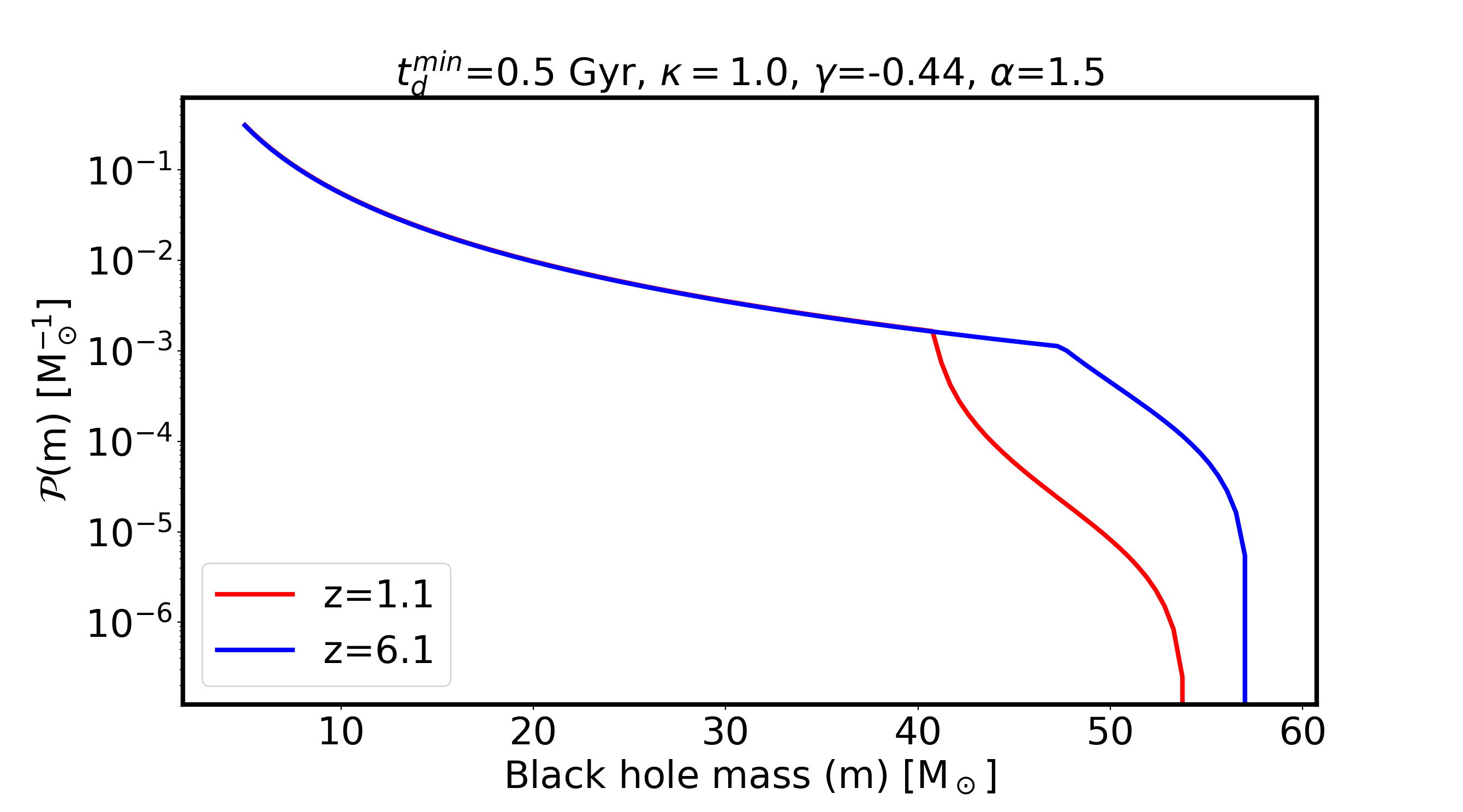}
     \caption{The probability distribution of the mass distribution is shown for two different redshifts.}
    \label{fig:probm}
\end{figure}

Fig. \ref{fig:window} shows that for different redshifts, the window function for the GW mass-scale is not the same at all redshifts. At higher redshifts, the PISN mass-scale exhibits a larger value of the PISN mass cut-off than at a lower redshift. This variation in the mass scale is more if the metallicity evolution with redshift is large or the delay time distribution is large or with a shallow evolution of the probability distribution ($\kappa$ close to zero). The shift in the PISN mass scale M$_{\rm PISN}$ and the maximum mass M$_{\rm max}$ with redshift is shown in Fig. \ref{fig:redshift}. It indicates that for different scenarios, there is a significant variation in the mass scales. This implies that the observed mass distribution of the BBHs is going to be significantly different because of the variation of the parameters related to metallicity and delay time distribution. For the scenarios considered in this analysis, the variation in PISN mass scale M$_{\rm PISN}$ can be between $10\%$ to $50\%$. The variation in the maximum mass scale M$_{\rm max}$ can be also around $25\%$. A previous study explored the metallicity dependence of the M$_{\rm max}$ \citep{2019ApJ...883L..24S} but did not show the impact of delay time on the mass distribution and how it can naturally produce a redshift dependence of the PISN mass scale and the smearing property of the BH mass distribution. The corresponding source frame probability mass distribution merging at a redshift $z_m$ can be  written as
\begin{equation}
    \mathcal{P}(m(z_m))= W(m(z_m))\mathcal{P}_s(m(z_m)),
\end{equation}
where $\mathcal{P}_s(m(z_m))$ is the probability distribution of the stellar compact objects that source the BH formation. This probability distribution can be considered to be motivated by a simple power-law form $m^{-\alpha}$. We assume in this paper that $\mathcal{P}_s(m(z_m))$ is redshift independent. But a redshift-dependent scenario is not outside the theoretical scope. The observed masses of the GW sources merging at redshift $z_m$ will be redshifted which can be written as
\begin{equation}
    m_{\rm det}= (1+z_m)m.
\end{equation}
We show the probability distribution of the masses, for this simple power-law form in Fig. \ref{fig:probm} for two different redshift values $z=1.1$ and $z=6.1$. So, even though the mass distribution of the BHs is extended to higher masses, the mass window function $\mathcal{W}(m(z_m))$ suppresses the masses above the mass scales set by the window functions at those two redshifts. The mass spectrum shows that the cutoff happens at different mass values at different redshifts. As result, the observed mass distribution of BHs is no more redshift independent. 

In summary, \textit{even a marginal dependence of the PISN mass scale of a BHs can lead to a much broad distribution of the PISN mass scale due to the mixing between the BHs formed at a different cosmic time due to a non-zero value of the delay time between the formation of a star and merger of a black hole. The probability distribution of the delay time leads to redshift dependence in the observed PISN mass scale and the maximum mass of a BH.}
We summarise below the key aspects of the redshift dependence of the mass distribution.

\textbf{The key aspects of this mass modeling}
\begin{itemize}
    \item This model predicts one of the potential sources of redshift dependence of the PISN mass scale M$_{\rm PISN}$ and shows how it is correlated with the quantities such as stellar metallicity, delay time distribution, and the value of the minimum delay time.
    \item This model naturally predicts a smearing behavior between the PISN mass scale and the maximum value of the mass distribution and how it evolves with redshift. 
    \item This model predicts a redshift dependence of the maximum mass value denoted by M$_{\rm max}$ and also its correlation with the PISN mass scale M$_{\rm PISN}$. The maximum mass scale also depends on the metallicity evolution with redshift and on the delay time distribution. 
    
\end{itemize}
 It is worth mentioning that the delay time distribution also impacts the redshift evolution of the merger rate of the BBHs \citep{2012ApJ...759...52D, Dominik:2014yma, Mapelli:2017hqk,2018MNRAS.474.2959G,Vitale:2018yhm,Toffano:2019ekp, Fishbach:2021mhp, Mukherjee:2021bmw}. So, joint estimation of the parameters related to the delay time distribution and the cosmological parameters can be made using their mass distribution and their distribution in luminosity distance.
 
 In light of the recent results from the LVK observations \cite{LIGOScientific:2018jsj, LIGOScientific:2020iuh,LIGOScientific:2021djp}, we have detected a few events whose masses are higher than the usual PISN scale considered as 45 M$_\odot$. As we have shown in Fig. \ref{fig:window}, due to a non-zero value of the delay time and evolution of the stellar metallicity, the mass distribution of the BHs can go up to high redshift. The window function $\mathcal{W}(m)$ gets a slope in its distribution, instead of a sharp cutoff. The slope of the mass window function can go beyond the value $45$ M$_\odot$, which can allow for heavier masses of the binary mergers. 
 High mass systems can be also be possible from several astrophysical scenarios as shown previously \citep{DiCarlo:2019fcq, Farrell:2020zju, Costa:2022aka, Briel:2022cfl}.  It is also worth mentioning that apart from the first-generation BHs, second-generation BHs due to hierarchical mergers can lead to heavier BHs. This can lead to additional variation in the BH mass distribution \citep{2021MNRAS.502.2049L,2022ApJ...929L...1B}. In summary, all these astrophysical uncertainties can make the mass distribution of the BHs redshift dependent which can make it more difficult to use for the cosmological purpose.
 
\section{Impact on the mass distribution of the GW sources}\label{redi}
The redshift evolution of the PISN mass scale with redshift will lead to a change in the mass distribution of BBHs with redshift. We study the impact on the observed mass distribution of the binary sources for two different network configurations, the current generation detectors such as LIGO \citep{LIGOScientific:2014pky}, Virgo \citep{Acernese_2014}, and the upcoming GW detectors such as KAGRA \citep{KAGRA:2020tym} and LIGO-India \citep{Unnikrishnan:2013qwa}, and the third generation GW detectors such as  Cosmic Explorer  \citep{Reitze:2019iox,2019CQGra..36v5002H}, and Einstein Telescope \citep{Punturo:2010zz}. The sources detectable from the current generation detectors and the next generation detectors will have a redshift reach up to $z=1$ and $z\sim 80$ respectively. In our analysis, we primarily focus on the GW sources for redshift up to $z=7$ for the third-generation detectors, since the expected merger rate at higher redshifts is not well known and difficult to model for the astrophysical BHs which are motivated by the Madau-Dickinson star formation rate \citep{Madau2014}. 

We sample GW sources with a mass distribution of the power-law form $m^{-\alpha}$ with $\alpha=2.35$ and a cut-off at a mass scale according to the mass window function which is a function of redshift. Along with a power-law form, we consider a bump at the M$_{\rm PISN}$ scale with the relative height of the peak in comparison with the low mass value as $\Lambda_g=0.1$, which is in the agreement with the GWTC-3 observation from LVK collaboration
\citep{Abbott_2021, LIGOScientific:2021psn,LIGOScientific:2021djp,Virgo:2021bbr} and also with the previous population synthesis models \citep{2019ApJ...882..121S}. The sources at higher redshifts have a mass cutoff at a higher value resulting in a broader mass spectrum at a high redshift than at a low redshift. We use a network of four GW detectors (LIGO-Hanford, LIGO-Livingston, Virgo, and KAGRA) to estimate the posteriors on the mass samples for GW sources distributed up to redshift $z=1.1$ which is possible to access from the current generation detectors \citep{KAGRA:2013pob}. 

\begin{figure}
    \centering
    \includegraphics[trim={0.cm 0.cm 0.cm 0.cm},clip,width=0.45\textwidth]{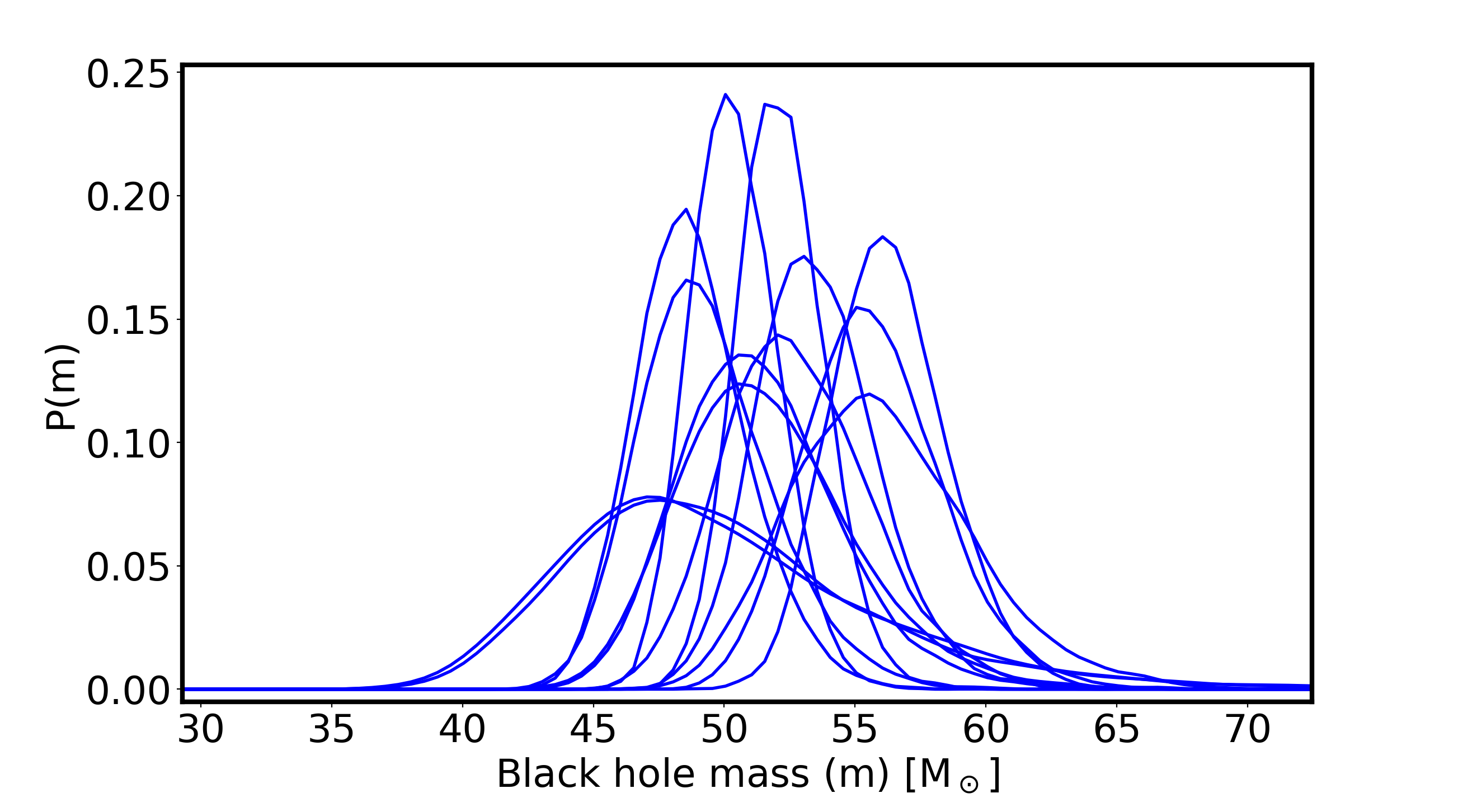}
    \includegraphics[trim={0.cm 0.cm 0.cm 0.cm},clip,width=0.45\textwidth]{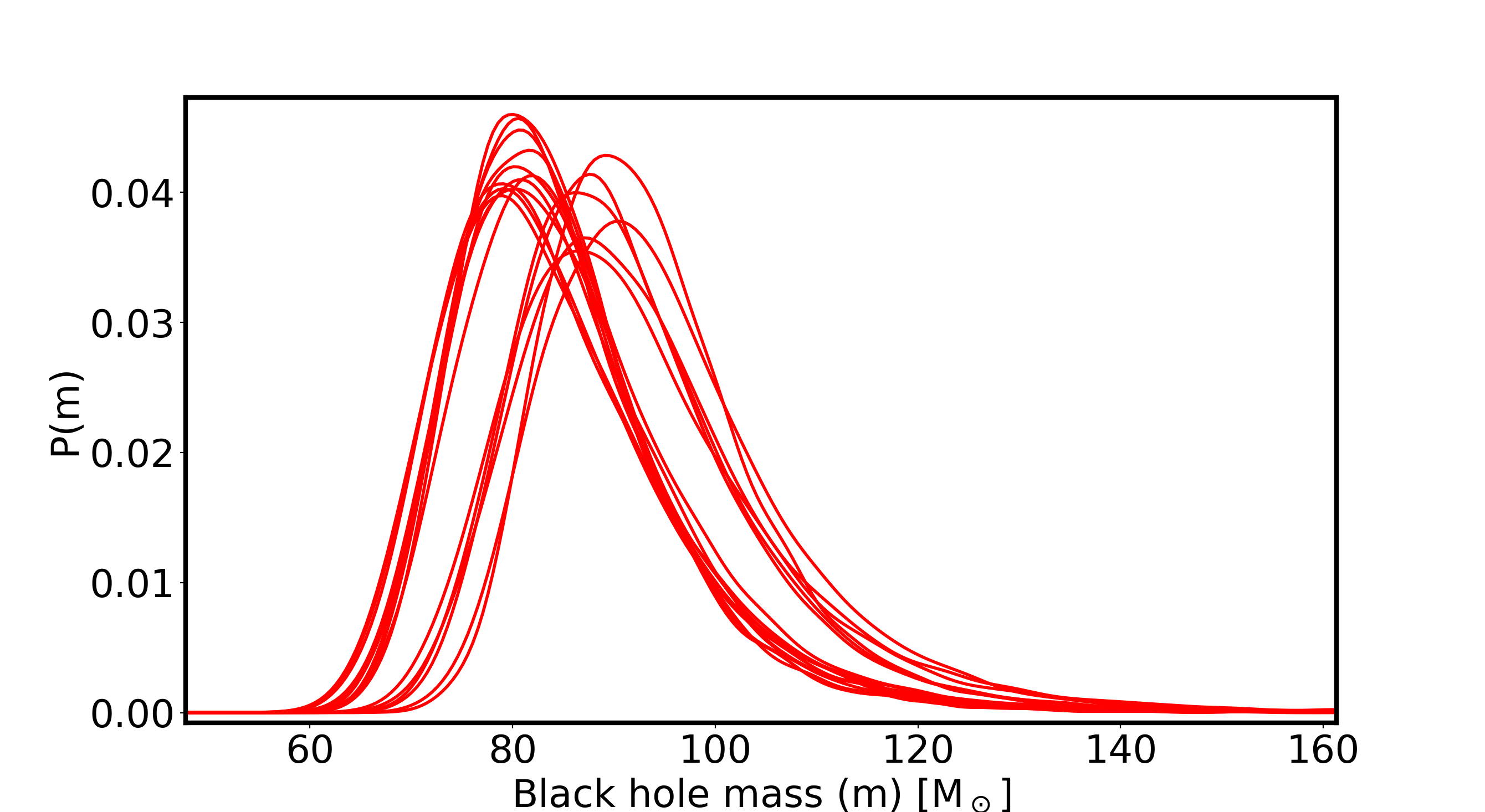}
      \captionof{figure}{The posterior distribution of the component masses for the objects close to the PISN mass are shown for redshift $z=0.2$ (blue) and redshift $z=1.0$ (red) for the LVK network of detectors with 2.5 years of observation time.}
    \label{fig:mass}
\end{figure}

We use an approximated form of the likelihood to infer the posteriors on the GW source parameter, following the previous analysis \citep{Farr:2019twy, Mastrogiovanni:2021wsd}. This assumption will not change the main conclusion of the paper on the incorrect estimation of the redshift using the PISN mass scale. The matched filtering signal to noise ratio (SNR), $\rho$ is estimated for BBHs with detector-frame chirp mass $\mathcal{M}_d$ at the luminosity distance of $d_\ell$, using the relation \citep{Farr:2015lna, 2021PhRvD.104f2009M}
\begin{equation}\label{snr}
    \rho= \rho_*\Theta\bigg(\frac{\mathcal{M}}{\mathcal{M}^*_d}\bigg)^{5/6}\bigg(\frac{d_\ell^*}{d_\ell}\bigg),
\end{equation}
where $\Theta$ is the detector projection factor that we assume as uniform between $[0,1]$. We have made a pessimistic choice of the projection factor to show the impact of the bias in inferring the redshift using GW sources. For sources, with a higher value of the projection factor, the bias in the redshift estimation can be even more noticeable.  The value of $\rho_*=8$ is set for the parameters $\mathcal{M}_d^*$ and $d_\ell^*$ for the LVK design sensitivity \citep{KAGRA:2013pob} and CE \citep{2019CQGra..36v5002H}. 
The network matched filtering  SNR is obtained by using $\rho^2_{\rm det}= \sum_i \rho_i^2$. We choose BBHs having $\rho_{\rm det} \geq 10$ as detected events. The estimated mass distribution as a function of redshift shows the variation due to the evolving mass distribution. This indicates that the assembling of the heavier objects takes place at higher masses for higher redshifts. The corresponding redshift evolution of the mass distribution is shown in Fig. \ref{fig:mass} only for the events which are close to the PISN mass scale.  

\section{Inference of the source redshift using the PISN mass cutoff}\label{inference}
The PISN mass scale can be used to identify the redshift of the GW sources if the mass of the PISN mass scale is redshift independent. However, as we have discussed in the previous sections, the PISN mass scale is going to vary with redshift. As a result, if we would like to infer the redshift of the GW sources for a fixed value of the PISN mass scale, then there is going to be a systematic error in the redshift estimation. The redshift dependence of the PISN mass scale depends on several astrophysical parameters (discussed above) and cannot be characterized uniquely.  

\begin{figure*}
    \centering
    \includegraphics[trim={0.cm 0.cm 0.cm 0.cm},clip,width=1.\textwidth]{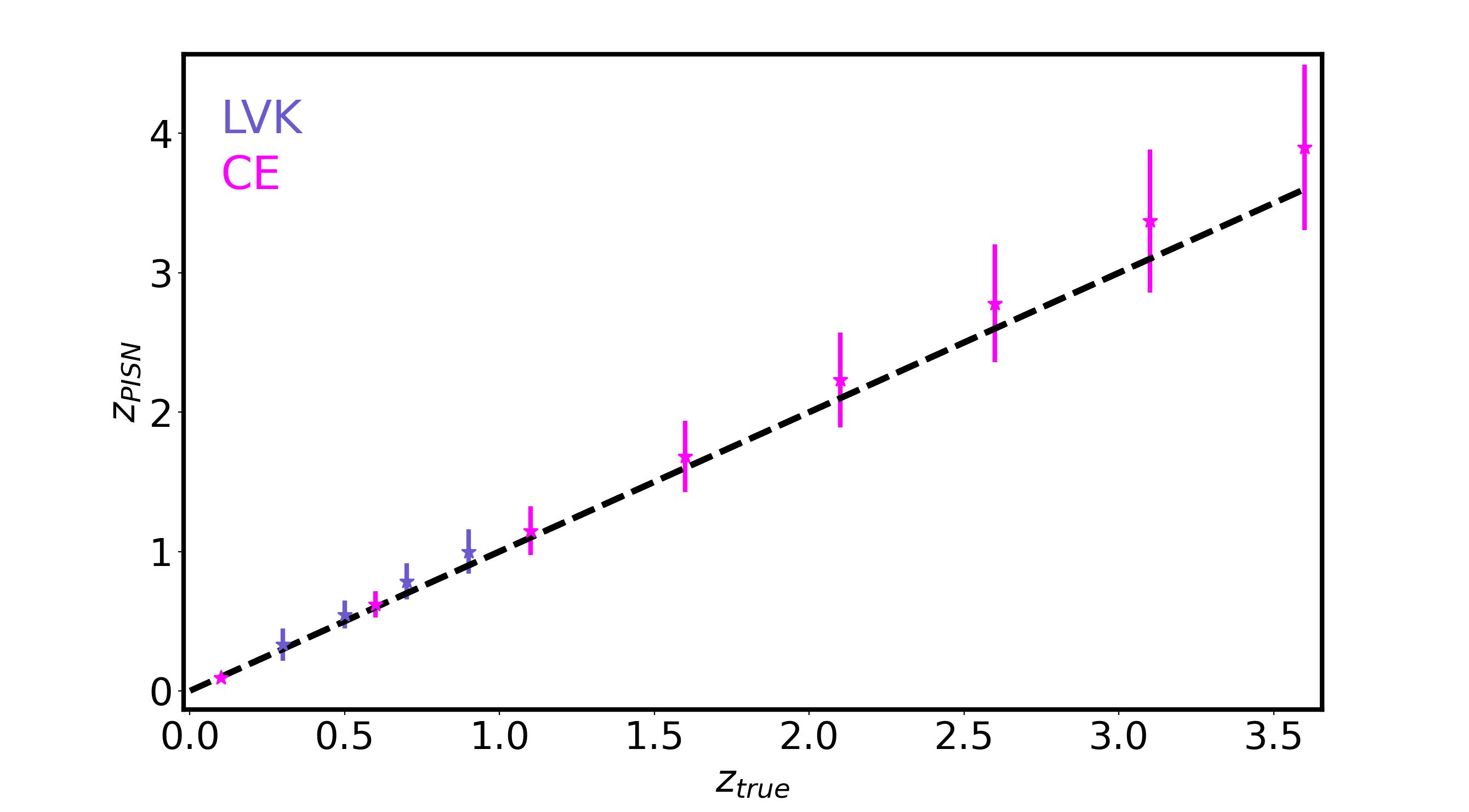}
    \includegraphics[trim={0.cm 0.cm 0.cm 0.cm},clip,width=1.\textwidth]{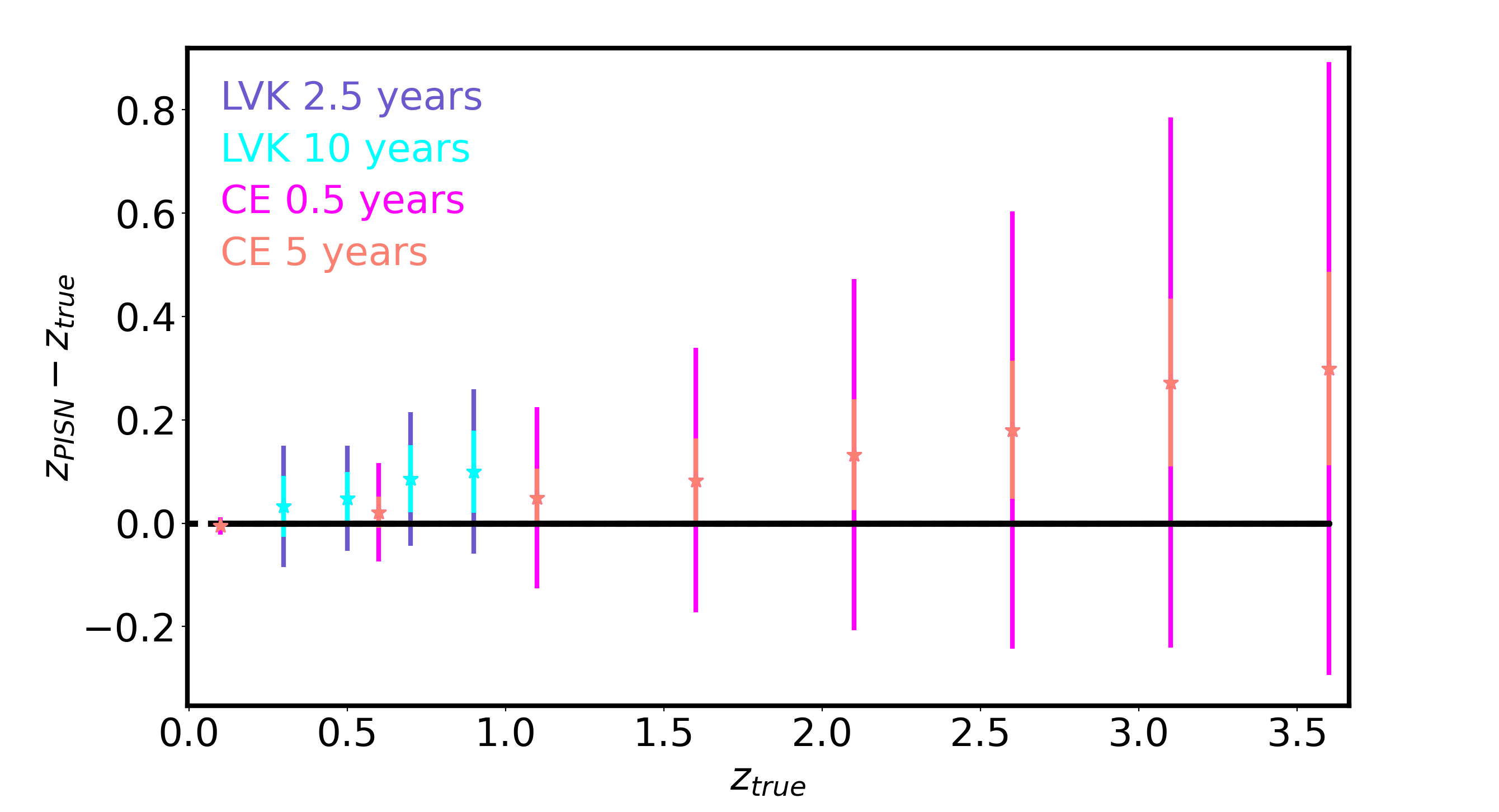}
     \caption{Top: The inferred redshift z$_{\rm PISN}$ assuming a fixed PISN mass scale with respect to the true redshifts z$_{true}$ of the samples is shown for LIGO-Virgo-KAGRA (LVK) and one Cosmic Explorer (CE) detector with 2.5 years and 0.5 years of observation time. Bottom: The difference between the inferred redshift z$_{\rm PISN}$ and the true redshift z$_{true}$ is shown for LVK with 2.5 years and 10 years and for one CE with 0.5 years and 5 years of observation time.} 
    \label{fig:redshifter}
\end{figure*}

We show the inferred redshift of the value assuming a fixed PISN mass scale indicated by z$_{\rm PISN}$ and the true redshift of the sources in the mock in Fig. \ref{fig:redshifter} along with the uncertainty due to the inferred mass distribution for 2.5 years of LVK observations and 0.5 years of Cosmic Explorer for the case with $t^{min}_d=0.5$ Gyr, $\kappa=1$, $\gamma=-0.44$, $\alpha=1.5$. This scenario is chosen as it has the minimum shift in the PISN mass scale in comparison to the other cases. So, it will describe the minimum error for the scenarios considered in this analysis. 
The deviation is much larger for the third-generation detectors such as Cosmic Explorer and Einstein Telescope than for the current generation GW detectors such as LVK. The systematic error in the inferred redshift is going to vary for different choices of the astrophysical parameters $t^{min}_d$, $\kappa$, $\gamma$, $\alpha$ considered in this analysis. For the cases with large variation in the M$_{\rm PISN}$ (see Fig. \ref{fig:redshift}), the corresponding systematic error in the inferred redshift z$_{\rm PISN}$ obtained using assuming a fixed of M$_{\rm PISN}$ will be larger than the case shown in Fig. \ref{fig:redshifter}. The difference between the inferred redshift and true redshift is shown in Fig. \ref{fig:redshifter} bottom panel for four different cases. The error bars on the redshift are the $68\%$ C.I. obtained from the inferred redshift by combining the mass posteriors. This indicates that with the increase in more events, the statistical error in the mass inference will be subdominant than the systematic error due to the unknown redshift dependence of the PISN mass scale. We find that there is going to be about $10\%$ variation for this model parameters. The variation can go as large as $25-30\%$ for different values of the parameters $t^{min}_d$, $\kappa$, $\gamma$, $\alpha$ considered in this analysis (see Fig. \ref{fig:window}). Moreover, this systematic uncertainty is only a minimum uncertainty. Incorrect astrophysical modeling, variation of metallicity with galaxies, multiple delay time distributions, and inaccurate prescriptions of stellar winds can increase this even further.  

\section{Impact on GW source population inference and cosmology}\label{cosmo}
From the above discussion, it is evident that the redshift inference from the PISN mass scale is not accurate and cannot be well predicted due to two major sources of uncertainties (i) redshift dependence of the metallicity and (ii) redshift dependence of the delay time distribution. Even in the scenario when the PISN mass scale can be correctly modeled \footnote{This is unlikely the case, as there exist several other unknown factors that can change the PISN mass scale.}, then also the observed mass population will not be redshift independent. This will have a significant impact on the inference of the GW source population as well as on the inference of cosmological results. 

\textit{Impact on the GW source population :} The redshift dependence of the GW mass distribution will have a significant impact on understanding the true mass distribution of the sources. Our study shows that the position of the PISN mass cutoff and the mass range over which the distribution can be smeared both are going to depend on redshift through the delay time distribution and metallicity. As a result, it is important to include the redshift dependence of the mass distribution of the GW sources in the analysis. Moreover, the redshift dependence of the mass distribution of the BBHs and also their merger rates are going to be correlated due to their dependence on the delay time distribution. So, a joint analysis of the GW mass distribution and the merger rates will be appropriate to infer the PISN mass scale, metallicity dependence, and the delay time distribution. In a recent work \citep{Karathanasis:2022rtr}, we made a joint estimation of these quantities from the GWTC-3 data of the LVK collaboration. An independent way to infer the PISN mass distribution is using the cross-correlation technique after marginalizing the GW bias parameters \citep{Mukherjee:2018ebj,Mukherjee:2019wcg,Mukherjee:2020hyn, Diaz:2021pem}. A proper inference of the mass distribution of the BBHs will also be useful for the analysis of the stochastic GW background \citep{Mukherjee:2019oma, Callister:2020arv, Mukherjee:2021ags, Mukherjee:2021itf} and lensing event rates \citep{Oguri:2018muv,Mukherjee:2021qam}. The amplitude of the stochastic GW background and also the lensing event rates strongly depend on the mass distribution of the BBHs. Inclusion of the redshift dependence of the PISN mass distribution explored in this analysis will have a vital role in correctly estimating these signals from the data. Moreover, constraints on the lensing event rates using the stochastic GW background \citep{Mukherjee:2020tvr,2020PhRvL.125n1102B} also need to use a mass distribution including the redshift dependence of the PISN mass-scale and the maximum mass. 

\begin{figure*}
    \centering
    \includegraphics[trim={0.cm 0.cm 0.cm 0.cm},clip,width=1.\textwidth]{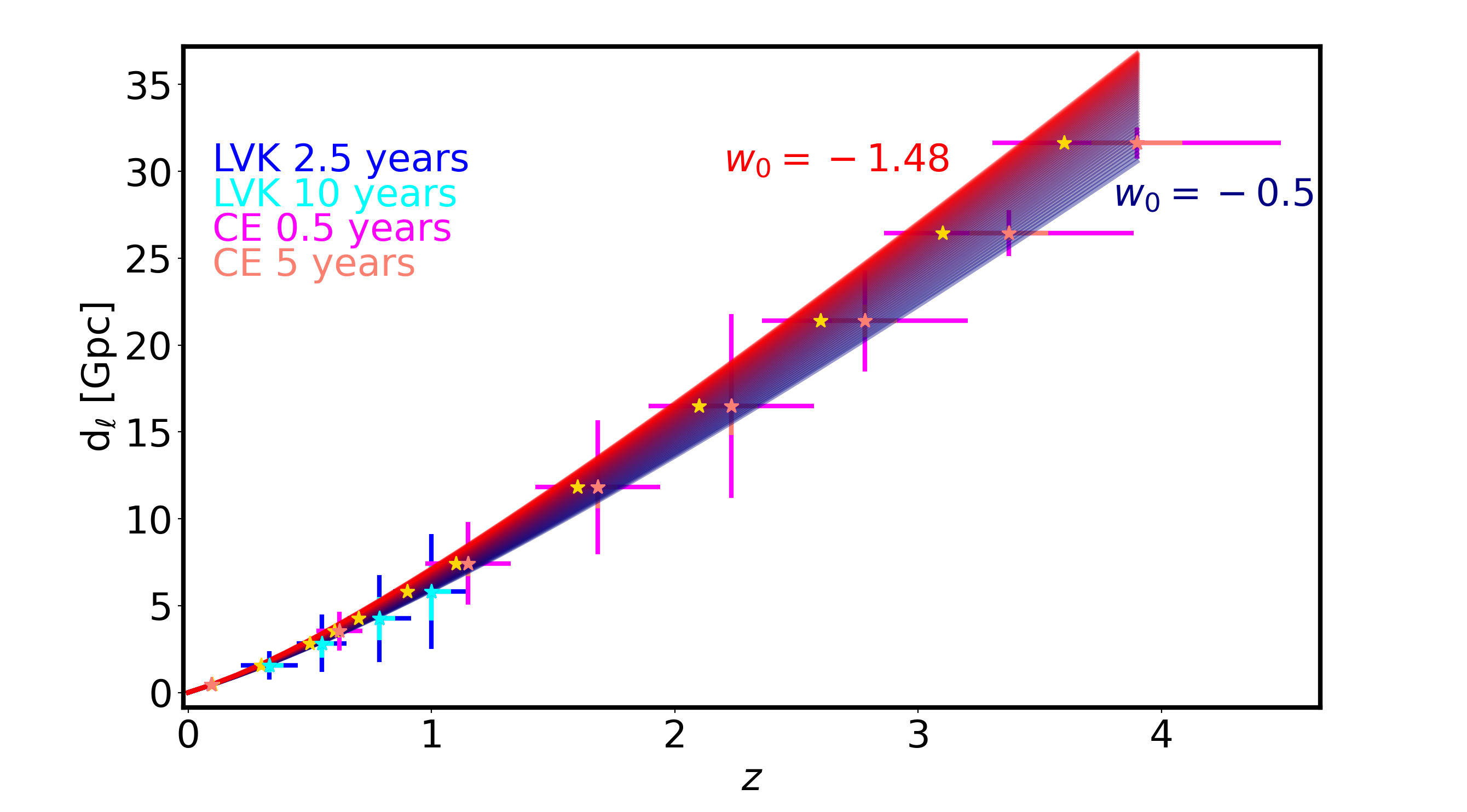}
     \caption{The impact of the incorrect estimation of the redshift on the luminosity distance $d_\ell$ and redshift $z$ plane is shown for LIGO-Virgo-KAGRA (LVK) for 2.5 years and 10 years and Cosmic Explorer (CE) for 0.5 years and 5 years along with the case for the true value of redshift in yellow. We show also a family of curves by changing only the dark energy equation of state parameter $w_0$ from -0.5 to -1.48. This indicates that an incorrect inference of the redshift from the PISN mass scale can lead to a map of an incorrect cosmic expansion history.} 
    \label{fig:dl}
\end{figure*}

\textit{Impact on the cosmological results :} The redshift dependence of the PISN mass cutoff leads to a systematic bias in the inference of the true cosmological redshift. As shown in the previous section, the bias in the redshift estimation is mainly more severe at the high redshift than at the low redshift due to the dependence on metallicity. As shown in Fig. \ref{fig:redshifter}, the error in the redshift estimation can be about $10\%$. This error can be as large as 30\% for the values of the parameters ($t^{min}_d$, $\kappa$, $\gamma$, $\alpha$) considered in this analysis. This error is going to be subdominant from the statistical error for the current LVK data GWTC-3 \citep{Abbott_2021, LIGOScientific:2021psn,LIGOScientific:2021djp,Virgo:2021bbr}, but in the future to achieve a few percent measures of the cosmological parameters such as the dark energy equation of state \citep{Farr:2019twy,You:2020wju}, the systematic uncertainty from the PISN mass cutoff will be a bottleneck. In Fig. \ref{fig:dl} we show the impact of the incorrect redshift inference on the luminosity distance ($d_\ell$) and redshift plane for LVK and CE, assuming the value of M$_{\rm PISN}$   as the one measured from low redshift ($z=0.2$). We also show the true injected value in the yellow marker. To compare this with different expansion history models, we vary a single parameter $w_0$ from -0.5 to -1.48, keeping all other cosmological parameters at a fixed value. This shows that for the inferred wrong redshifts using the PISN mass scale, there is going to be a significant bias in the value of inferred dark energy equation of state parameter towards a higher value ($w_0>-1$). This value is about a factor of two away from the true value. Similar to $w_0$, the systematic effect will also influence the measurement of the Hubble constant and other cosmological parameters that appears in the Hubble parameter $H(z)$. So, without carefully mitigating the systematic effect due to the PISN mass scale, one cannot use it for accurate inference of the cosmological parameters.  The errors shown in Fig. \ref{fig:redshifter} are for a scenario with a very simplistic model of the metallicity dependence and delay time distribution. The ignorance of the true values of these quantities leads to a systematic overestimate of the redshift of the GW sources. To reach a few percent level measurements of the cosmological parameters using the PISN mass cutoff, we need to mitigate this systematic error, otherwise, it will lead to a systematic bias in the inference of the cosmological parameters. Careful calibration of the mass distribution of BBH is required in order to mitigate possible systematic effects.    

Similar to the background cosmological parameters, the use of the PISN mass cutoff was also thought to be useful to test the general theory of relativity (GR) from GW propagation by inferring the frictional term \citep{Belgacem:2017ihm, Belgacem:2018lbp,Ezquiaga:2021ayr,Mancarella:2021ecn} and also for measuring extra dimensions \citep{Hernandez:2021qfp}. The measurement of the deviation from the GR depends completely on the difference in the luminosity distance inferred from GW and the luminosity distance for photons at that redshift. Any systematic error in the redshift inference for the GW sources will lead to a bias in the value of frictional terms. For most of the non-GR theories, the deviation from GR is expected of the order of a few percent. As a result, to make a robust measurement of any deviation from GR, it will be essential to correctly include the systematic error that can be induced from the redshift dependence of the PISN mass cutoff. Our analysis shows that there can be a 10-30\% systematic error in PISN mass scale over the redshift up to $z=7$. As a result, a few percent level measurement is not possible unless this effect can be properly mitigated. Several recent analyses \citep{Ezquiaga:2021ayr,Mancarella:2021ecn,Hernandez:2021qfp} assume a fixed PISN mass scale and hence these measurements are not reliable and subject to systematic errors. 
The inference of the cosmological parameters jointly with the frictional term from GW propagation can be made in a robust way using the cross-correlation technique from BBHs and inferring the clustering redshift \citep{Mukherjee:2020mha, 2021ApJ...918...20C}.

\section{conclusion and the way forward}\label{conc}
In this work, we explore for the first time the redshift dependence of the PISN mass cutoff and its implication on the estimation of the source redshift of BBHs. We show that even if the PISN mass cutoff and its dependence on the stellar metallicity can be well modeled theoretically, the redshift evolution of the metallicity and the delay time distribution of the BBH mergers can lead to a mixing between BHs formed across a large cosmic epoch. This will lead to redshift dependence of the PISN mass cutoff. The unknown form of the probability distribution of the delay time will lead to an unknown amount of systematic shift in the PISN mass cutoff to higher values with a change in the redshift. The mass scales over which the BH distribution will be smeared to zero also depend on the metallicity and the delay time distribution. Ignoring the redshift dependence for any cosmological and GW population analysis can lead to a systematic bias. 

We show that the use of a fixed PISN mass cutoff to infer the source redshift of the BBHs can lead to a biased inference. Even for one of the best cases (minimum shift in the PISN mass cutoff) and assuming that one can model the metallicity dependence of the PISN mass cutoff theoretically, we find that there is going to be around $10-30\%$ systematic error in the inference of the redshift of the sources (towards high values). Though this error is sub-dominant to the statistical error in the mass measurement of the current LVK data \citep{Abbott_2021, LIGOScientific:2021psn,LIGOScientific:2021djp,Virgo:2021bbr} and the cosmological results inferred from GWTC-3 \citep{Virgo:2021bbr} are unlikely to be influenced by this, but it is going to be a major source of uncertainty in the future when statistical error will be sub-dominant. This is also in agreement with the recent studies from GWTC-2 and GWTC-3 \citep{2021ApJ...912...98F, LIGOScientific:2021psn}. We show that for a network of four GW detectors such as LIGO-Hanford, LIGO-Livingston, Virgo, and KAGRA and also in the future from Cosmic Explorer, the systematic error will lead to a systematic bias in the inferred value of redshift at the level of 10-30\%. As a result, to reach a percent level measurement of the cosmological parameter using the PISN mass cutoff, it is essential to mitigate this systematic error. Similarly, a wrong inference of the redshift will have an impact on the measurement of any non-GR signatures due to GW propagation. 

The redshift dependence of the PISN mass cutoff modeled in this paper will also be useful for understanding the source frame mass distribution of the BBHs. The inclusion of a redshift dependence of the PISN mass cutoff and exploring its dependence on the stellar metallicity and delay time distribution will able to shed light on the mass distribution of the BHs in the Universe. A systematic shift in the PISN mass cutoff with redshift and its dependence on the delay time will be able to help us in understanding the formation channel of the BBHs. An independent way to infer the redshift dependence of the PISN mass scale is by inferring the clustering redshift using the cross-correlation technique and marginalizing over the GW bias parameters \citep{Oguri:2018muv,Mukherjee:2018ebj,Mukherjee:2020hyn, Scelfo:2020jyw,Diaz:2021pem}.

In summary, this work shows for the first time the redshift independence of the PISN mass cutoff is not possible in reality. Even when there is no theoretical modeling uncertainty, astrophysical uncertainties will play a significant role in inducing the redshift dependence of the PISN mass cutoff. In addition, there can be also secondary sources, and additional theoretical uncertainties that even increase the budget of the systematic error. As a result, the use of the PISN mass cutoff for inferring redshift to the source, and use that for cosmological analysis can cause a significant systematic bias. Though currently, the statistical error is larger than the systematic error, in the future, an accurate measurement of the cosmological parameters will not be possible unless this can be mitigated by other techniques. In future work, we will develop techniques to also include the redshift dependence of the PISN mass cutoff and its joint estimation with the cosmological parameters.  

\section*{Acknowledgements}
S.M. is grateful to Simone Mastrogiovanni for reviewing the manuscript as a part of the LSC review process and providing very useful suggestions. The author acknowledges useful discussions and feedback from Jonathan Gair, Martin Hendry, Gilbert Holder, Matthew Johnson, and Luis Lehner. S.M. thank Robert Farmer for pointing to a typographical error in the paper. 
S.M. is supported by the Simons Foundation. Research at Perimeter Institute is supported in part by the Government of Canada through the Department of Innovation, Science and Economic Development and by the Province of Ontario through the Ministry of Colleges and Universities. This analysis is carried out at the computing facility of the Perimeter Institute and the Infinity cluster hosted by Institut d'Astrophysique de Paris. We thank Stephane Rouberol for smoothly running the Infinity cluster. We acknowledge the use of following packages in this work: Astropy \citep{2013A&A...558A..33A, 2018AJ....156..123A}, IPython \citep{PER-GRA:2007}, Matplotlib \citep{Hunter:2007},  NumPy \citep{2011CSE....13b..22V}, and SciPy \citep{scipy}. The author would also like to thank the LIGO-Virgo-KAGRA scientific collaboration and Cosmic Explorer collaboration for providing the noise curves. LIGO is funded by the U.S. National Science Foundation. Virgo is funded by the French Centre National de Recherche Scientifique (CNRS), the Italian Istituto Nazionale della Fisica Nucleare (INFN), and the Dutch Nikhef, with contributions by Polish and Hungarian institutes. This material is based upon work supported by NSF's LIGO Laboratory which is a major facility fully funded by the National Science Foundation. 

 \section*{Data Availability}
The data underlying this article will be shared at request to the corresponding author. 

\bibliographystyle{mnras}
\bibliography{main_MNRAS}
\label{lastpage}
\end{document}